# Evaluating Load Models and Their Impacts on Power Transfer Limits

Xinan Wang, *Student Member, IEEE*, Yishen Wang, *Member, IEEE*, Di Shi, *Senior Member, IEEE*,
Jianhui Wang, *Senior Member, IEEE*, Siqi Wang, *Member, IEEE*, Ruisheng Diao, *Senior Member, IEEE*,
Zhiwei Wang, *Senior Member, IEEE*

*Abstract*— Power transfer limits or transfer capability (TC) directly relate to the system operation and control as well as electricity markets. As a consequence, their assessment has to comply with static constraints, such as line thermal limits, and dynamic constraints, such as transient stability limits, voltage stability limits and small-signal stability limits. Since the load dynamics have substantial impacts on power system transient stability, load models are one critical factor that affects the power transfer limits. Currently, multiple load models have been proposed and adopted in the industry and academia, including the ZIP model, ZIP plus induction motor composite model (ZIP + IM) and WECC composite load model (WECC CLM). Each of them has its unique advantages, but their impacts on the power transfer limits are not yet adequately addressed. One existing challenge is fitting the high-order nonlinear models such as WECC CLM. In this study, we innovatively adopt double deep Q-learning Network (DDQN) agent as a general load modeling tool in the dynamic assessment procedure and fit the same transient field measurements into different load models. A comprehensive evaluation is then conducted to quantify the load models' impacts on the power transfer limits. The simulation environment is the IEEE-39 bus system constructed in Transient Security Assessment Tool (TSAT).

## I. Background

The power transfer limit or transfer capability (TC) refers to the maximum power that can be safely transferred between a generator group and a load center [1], in which the generator group is called the power source and the load center is called the power sink. The power transfer limit is constrained by both static constraints, such as transmission line thermal limits, and dynamic constraints, such as transient stability limits, voltage stability limits and small-signal stability limits [2]. In this paper, we evaluate the power transfer limit from a transient stability perspective, more specifically, the load model's impacts on the transfer limits.

From the mathematical point of view, the power system transient stability analysis (TSA) is a large-scale nonlinear and non-convex problem which contains differential algebraic equations [3]. In order to calculate the proper power transfer limits, the system operators need to adopt a load model first and screen every possible contingency by repeatedly running the TSA program. To avoid the overestimation of the limit, besides the major constraints mentioned earlier, a variety of supplemental static and dynamic constraints have been considered in the calculation. For instance, in [1], the authors included the electro-thermal coupling effect of the transmission line into the calculation, and concluded that the conventional transmission line thermal analysis method overestimates the power transfer limit. In [4], the authors considered the thermal balance of the overhead conductor in the analysis and indicated that the ambient temperature and line length can significantly influence the TC calculation. Dynamics in the power system also play an important role in the transfer limit analysis, which mainly refers to the transient dynamics of generators and load models. The influence of the generator dynamics on the power transfer limit has been widely studied. In [2], the authors conducted a trajectory sensitivity analysis on the dynamic components in the generator and develop an accurate transfer limit assessment method with much fewer computation efforts compared with conventional methods. In [5], the authors suggested that if the reactive power limit of a generator is encountered, the system stability can be severely harmed. In recent years, renewable penetration keeps increasing, which changes the system dynamic characteristics. Therefore, in [6], the authors investigated the PV plants' impacts on the transfer limit and developed an adaptive reactive power droop control scheme to mitigate these impacts.

However, there is no complete study to analyze and compare the influence of load model dynamics on TC. In [8] - [9], the authors only examined the static load models' impact on the transfer limit, such as constant impedance, constant current, constant power, and exponential load model. In practice, static load models are also dominant [10]. Only about 30% of system operators consider dynamic loads in their load models, which is partly because of the difficulties in fitting the field measurements into the non-linear and non-convex differential algebraic equations of dynamic loads. Enabled by the reinforcement-learning technique development in power systems [11], a two-stage reinforcement learning-based load modeling method [12] is developed and can overcome the mentioned difficulties and make it possible to even accurately fit the state-of-the-art WECC composite load model, which contains more than 130 parameters.

In this paper, we apply a universal load modeling method which is based on a double deep Q-learning network (DDQN) on load models such as ZIP, ZIP + IM, and WECC CLM. By fitting the same event record into those models, their impacts

This work is funded by SGCC Science and Technology Program, "Research on Real-time Autonomous Control Strategies for Power Grid based on AI Technologies", under contract no. 5700-201958523A-0-0-00. X. Wang, Y. Wang, D. Shi, S. Wang, R. Diao and Z. Wang are with GEIRI North America, San Jose, CA 95134, USA. X. Wang and J. Wang are with the Department of Electrical and Computer Engineering, Southern Methodist University, Dallas, TX 75205 USA.

on the transfer limits are comprehensively evaluated. Several observations are made based on the assessment results to provide further discussions. The transfer limit assessment is conducted on the IEEE 39-bus system using commercial power system simulation software Transient Security Assessment Tool (TSAT). To the best of our knowledge, it is the first paper that proposes a general deep reinforcement learning based load modeling methods that can be applied to different load models, which significantly decreases the complexities of comparing the system transfer limits under different load models.

The remainder of this paper is organized as follows. Section II introduces the load models considered in this study. In section III, the reinforcement learning-based load modeling setup for different load model is introduced. In section IV, case studies and concluding remarks are provided.

## II. INTRODUCTION OF LOAD MODELS

### A. ZIP Static Load Model

The ZIP model is a widely accepted static model in both academia and industry; it has fewer parameters and is capable of simulating a variety of dynamics in the powers system [13]. As shown in Figure 1, in the ZIP model, Z represents a constant impedance load, I refer to constant current load, and P denotes the constant power load. The real and reactive power of a ZIP load is a second-order polynomial function of the transient voltage, which is shown from (1) – (3):

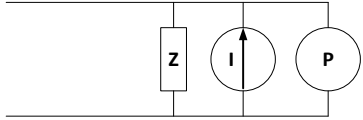

Figure 1. Structure of a ZIP load

$$p_{zip} = p_{0,zip} \cdot (p_{1c} \cdot \left(\frac{V}{V_0}\right)^2 + p_{2c} \cdot \frac{V}{V_0} + p_{3c}) \quad (1)$$

$$q_{zip} = q_{0,zip} \cdot (q_{1c} \cdot \left(\frac{V}{V_0}\right)^2 + q_{2c} \cdot \frac{V}{V_0} + q_{3c}) \quad (2)$$

$$\begin{cases} p_{1c} + p_{2c} + p_{3c} = 1, \ 0 \leq p_{1c}, p_{2c}, p_{3c} \leq 1 \\ q_{1c} + q_{2c} + q_{3c} = 1, \ 0 \leq q_{1c}, q_{2c}, q_{3c} \leq 1 \end{cases} \quad (3)$$

Where $p_{0,zip}$ and $q_{0,zip}$ are the steady-state active and reactive power taken by the ZIP load. $p_{1c}, p_{2c}, p_{3c}$ and $q_{1c}, q_{2c}, q_{3c}$ are the ratios of real and reactive power consumed by constant impedance, constant current, and constant power load in total real and reactive load. Some system operators simplify the ZIP load model by only using one or two components. According to the survey in [10], 13% of the system operators use constant P as their load models, and 13% of them use constant I as their load models.

### B. Induction Motor (IM)

The squirrel-cage induction motor is typically used to represent the dynamic load in transient stability analysis [14]. The simplified electrical circuit for the IM is shown in Figure 2. The active/reactive power consumptions of the IM are shown from (4) – (8):

$$\frac{dv'_{d,t}}{dt} = \frac{-R_r}{X_r+X_m}\left(v'_{d,t} + \frac{X_m^2}{X_r+X_m}i_{q,t}\right) + s_t \cdot v'_{q,t} \quad (4)$$

$$\frac{dv'_{q,t}}{dt} = \frac{-R_r}{X_r+X_m}\left(v'_{q,t} - \frac{X_m^2}{X_r+X_m}i_{d,t}\right) - s_t \cdot v'_{d,t} \quad (5)$$

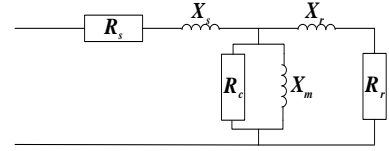

Figure 2. Equivalent circuit for an induction motor

$$\frac{ds_t}{dt} = \frac{1}{2H}[T_{m0}(1-s_t)^2 - v'_{d,t}i_{d,t} - -v'_{q,t}i_{q,t}], \quad (6)$$

$$P_{IM,t} = [R_s(u_{d,t}^2 + u_{q,t}^2 - u_{d,t}v'_{d,t} - u_{q,t}v'_{q,t}) - X'(u_{d,t}v'_{q,t} - u_{q,t}v'_{d,t})]/(R_s^2 + X'^2) \quad (7)$$

$$Q_{IM,t} = [X'(u_{d,t}^2 + u_{q,t}^2 - u_{d,t}v'_{d,t} - u_{q,t}v'_{q,t}) - R_s(u_{d,t}v'_{q,t} - u_{q,t}v'_{d,t})]/(R_s^2 + X'^2) \quad (8)$$

Where, $R_s$ and $X_s$ are the stator resistance and reactance, $R_r$ and $X_r$ are the rotor resistance and reactance, $X'$ denotes the short circuit reactance, $v'_{d,t}$ and $v'_{q,t}$ represent the transient voltage for IM on d- and q-axis, $u_{d,t}$ and $u_{q,t}$ are the bus voltages on d- and q-axis, $i_{d,t}$ and $i_{q,t}$ represent the d/q-axis stator current. The ZIP+IM composite load model [15] is obtained by connecting the ZIP model in parallel with the IM model. One issue raised for ZIP+IM is not being able to simulate the fault-induced delayed-voltage-recovery (FIDVR) event [16], which is primarily caused by stalling of the single-phase motors (e.g., the residential AC compressors).

### C. WECC Composite Load Model

To better capture FIDVR events as well as other new types of dynamics that brought by distributed energy resources (DER) and loads interfaced via power electronics, the Western Electricity Coordinating Council (WECC) comes up with the WECC composite load model (WECC CLM). WECC CLM is mainly formed by substation, feeder, and load, as shown in Figure 3. The substation is simulated by a load-tap-changing (LTC) transformer and a shunt capacitor; the feeder is modeled by a conventional transmission line pi-model; at the end-user side, there is one composite load which includes three three-phase IMs: Ma, Mb, and Mc, one single-phase IM: Md, one ZIP load and one electronic load.

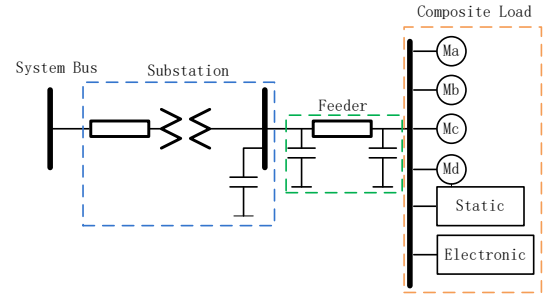

Figure 3. Structure of a WECC CLM

Different from the third-order IM model introduced previously, the IM model used in the WECC CLM is the fifth-order model, which considers the sub-transient voltage dynamics as well. Due to space limits, the detailed mathematic formulation of the fifth-order IM model is not presented in this paper but can be accessed from [12]. According to [17], Ma represents the three-phase IMs that drive constant torque loads, such as commercial/industrial air conditioner; Mb denotes the IMs of torque speed-squared loads with high inertia, such as fan motors used in residential and commercial buildings; Mc is the three-phase IMs with torque speed-squared loads with low inertia, such as direct-connected pump motors used in commercial buildings. The single-phase IM model is a performance-based model and developed based on extensive laboratory testing conducted by WECC, which includes the protection relays and the compressors in the residential AC units. The electronic load is modeled as a piece-wise linear function of the bus voltage with a unity power factor.

## III. REINFORCEMENT LEARNING-BASED LOAD MODELING CONTROL

As discussed earlier, many utilities only use static load models in their transfer limit analysis and neglect the impacts of dynamic loads. For these utilities that use dynamic load models, most of them choose the ZIP + IM model [19], which is recognized for not fully adequate to address the current load characteristic transitions [16]. Up to now, WECC CLM is widely accepted as the state-of-the-art load model [18] due to its robustness in modeling a variety of load compositions and its capability of simulating the electrical distance between the end-users and the transmission substations. However, the WECC CLM contains more than 130 parameters with complex control logics which make the conventional load modeling methods that work well on the ZIP + IM model no longer effective. The parameter settings for the WECC CLM are determined by load survey of empirical data [20], and users do not have a systematic method to identify the model parameters.

As discussed in [12], a DDQN agent essentially apply control actions on the load composition of the WECC CLM and effectively model the other parameters. We reformulate the DDQN-based control framework into a more general form and enable the DDQN agent to work on other load models, including ZIP, ZIP + IM and WECC CLM. By applying this method on different load models with respect to the same event data, we can approximate an aggregated load with different sets of parameters. An extensive evaluation of these load models' impacts on the transfer limits is then conducted.

### A. Reinforcement Learning-based Load Modeling Strategy

For this DDQN-based two-stage load modeling method, the DDQN agent works in stage one. In this stage, the objective of the agent is to find a load composition to represent the measured load dynamic responses. The state variable or the input to the agent is the current load fractions of all the load elements. The decision variables or control actions of the agent is the load fraction adjustment to be implemented among different load components. Without prior information on the other parameters in the load model, the agent will randomly sample their values within a defined range [20].

Due to the strong uncertainties on these sampled parameters, the load model's dynamic responses from a certain load composition contains large variances. Guided by a designed reward function, the DDQN agent will iteratively try and search for a load composition that has the highest probability to generate dynamic responses that are close to the reference dynamic responses. After the load composition is determined in stage one, a Monte-Carlo simulation is then conducted to estimate other load parameters for the purpose of a precise matching with the reference dynamic responses. Including WECC CLM, we reformulate this framework for other load models, such as ZIP, ZIP+IM, as shown in (12)-(14):

Constraints for WECC CLM model:

$$\begin{cases} P_{CLM} = P_{elc} + P_{zip} + P_{IMA} + P_{IMB} + P_{IMC} + P_{IMD} \\ Q_{CLM} = Q_{elc} + Q_{zip} + Q_{IMA} + Q_{IMB} + Q_{IMC} + Q_{IMD} \\ f_{elc} + f_{zip} + f_{IMA} + f_{IMB} + f_{IMC} + f_{IMD} = 1 \end{cases} \quad (12)$$

Constraints for ZIP + IM model:

$$\begin{cases} P_{ZIP+IM} = P_{ZIP} + P_{IM} \\ Q_{ZIP+IM} = Q_{ZIP} + Q_{IM} \\ f_{ZIP} + f_{IM} = 1 \end{cases} \quad (13)$$

Constraints for ZIP model:

$$\begin{cases} P_{zip} = P_{z,p} + P_{i,p} + P_{p,p} \\ Q_{zip} = Q_{z,q} + Q_{i,q} + Q_{p,q} \\ f_z + f_i + f_p = 1 \end{cases} \quad (14)$$

In (12), $P_{CLM}$ and $Q_{CLM}$ are the total active and reactive load for a WECC CLM load; $f_{elc}, f_{zip}, f_{IMA}, f_{IMB}, f_{IMC}$, and $f_{IMD}$ are the load fractions for electronic load, ZIP load, induction motor A, induction motor B, induction motor C and single-phase induction motor D. The summation of the those load fractions is one. Following the same manner, ZIP + IM load defines ZIP and induction motor components in (13) with only two load fractions. In (14), $P_{zip}$ and $Q_{zip}$ are the total active and reactive load for a ZIP load; $f_z, f_i$, and $f_p$ are the load fractions for constant impedance load $[p_{1c}, q_{1c}]$, constant current load $[p_{2c}, q_{2c}]$, and constant power load $[p_{3c}, q_{3c}]$. These three are vectors to characterize both active and reactive load fractions for this ZIP type, which are the decision variables to be optimized through the DDQN agent.

When fitting a dynamic load response into a model using the proposed method, the DDQN agent will take control actions to adjust the load fractions among different load components based on the current load composition. This means the state taken by the agent is the load fraction: $s = [f_1, f_2, ..., f_N]$, $(\sum_{k=1}^{N} f_k = 1)$; the actions taken by the agent is: $a = [\cdots, \Delta f, \cdots, -\Delta f, \cdots]$, where $\Delta f$ is the fraction modification value. Each action only has two non-zero elements $\Delta f$ and $-\Delta f$. The number of actions is $n \cdot (n-1)$ and $n$ refers to the number of components contained in the load model. For example, 6 actions for ZIP, 30 for WECC CLM.

After each action being taken, the previous state $s$ turns into a future state $s'$. The summation of the new state $s'$ remains one. This training process is summarized in Figure 4. The objective function for this fraction control problem is shown in (15), which minimizes the loss function to quantify the diffences between the dynamic responses generated by the fitted load model and the reference responses collected from the field measurement. Equation (16) is the customized loss used in this application, which contains $\mathscr{L}$-2 root mean squared error term and peak/valley index mismatch term. The agent will continue to apply control actions until the loss is less than a predefined threshold.

The $Q$ function for the DQN agent and its learning objective are shown in (17) and (18). In (14), $Q^A$ and $Q^B$ are the two $Q$ functions learned by the agent A, and Agent B. Agent B's parameters have a delayed update compared to the agent A, this training setup is called the double deep Q-learning network (DDQN) [21].

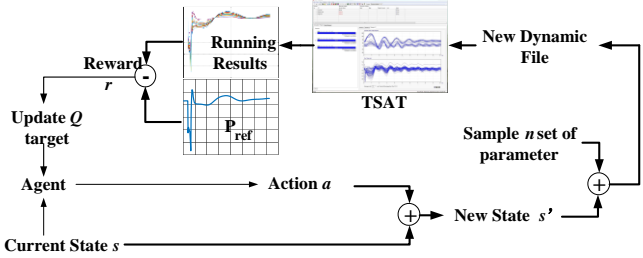

Figure 4. Reinforcement learning-based load modeling process

$$Obj: \min \frac{1}{n \cdot m} \sum_{j=1}^{m} \sum_{i=1}^{n} \mathcal{L}(\boldsymbol{P}_{fit,j}^i, \boldsymbol{P}_{ref,j}) + \mathcal{L}(\boldsymbol{Q}_{fit,j}^i, \boldsymbol{Q}_{ref,j}) \quad (15)$$

$$\mathcal{L}(\boldsymbol{A}, \boldsymbol{B}) = \alpha \frac{\sum_{j=1}^{L}(A_j - B_j)^2}{L} + \beta(|idx(A_{min}) - idx(B_{min})| + |idx(A_{max}) - idx(B_{max})|) \quad (16)$$

There are significant differences between fitting the static ZIP load, ZIP + IM loads and WECC CLM using the reinforcement learning method. For the ZIP, the variables are the ZIP load composition $[p_{1c}, p_{2c}, p_{3c}, q_{1c}, q_{2c}, q_{3c}]$, which is directly solved by the stage one, and stage two is not needed. But for ZIP + IM and WECC CLM, besides identifying the load composition between different load components, there are also parameters for dynamic loads and control settings that need to be identified. Then, in the training process, after each load composition adjustment, multiple sets of parameters will be uniformly sampled by the agent. Their average fitting error, as shown in (15), will be used as the reward to the agent. The exact values for these parameters will be determined in stage two.

$$Q^A(s, a) = (1-\alpha)Q^A(s,a) + \alpha \cdot (r + \gamma \cdot \max Q^B(s', a)) \quad (17)$$

$$\min(\mathcal{L}) = \|Q^A(s,a) - r - \gamma \cdot \max Q^B(s',a)\| \quad (18)$$

Due to the uncertainties contained in the load model parameters, the DDQN agent will offer multiple load component solutions in the stage one. To rank the probability of each candidate solution, a quantile loss function named pinball loss [23] is applied as shown in (19):

$$L_\tau(\hat{x}_o, x) = \max[(\hat{x}_o - x) \cdot \tau, (\hat{x}_o - x) \cdot (\tau - 1)] \quad (19)$$

In (19), $\hat{x}_o$ is the value at quantile $o$ of a group of data, $x$ indicates the value that needs to be evaluated, and $\tau$ refers to the penalize factor. By comparing the average pinball loss of all the snapshots in reference curves $P_{ref}$ and $Q_{ref}$ under each candidate load composition, we can select the one with the lowest losses.

### B. Transfer Limit Assessment

In our study, we investigate the impacts of load models on the transfer limit between any two zones of the test system. During the assessment, the active power consumed by a sink in one zone will be increased from its base load $P_{base}$ at a step of $\Delta P$ at each time. An N-1 contingency analysis will be conducted on the system to decide if the system violates the static limits and transient stability limits. Such an assessment process stops until the maximum power $P_{max}$ is found so that a $P_{max} + \Delta P$ load consumption at the sink can cause the system violating the static limits or transient stability limits. Then, $P_{max}$ is regarded as the transfer limit between the two zones.

To have an accurate transfer limit assessment, the load model at the sink is expected to have very similar dynamic responses with the historic event records. Therefore, it is of great importance to conduct a load modeling and parameter identification before the transfer limit assessment. Following the formulation mentioned from (12)-(14), we train multiple DDQN agents in parallel with each agent corresponding to one load model. Then, we apply these fitted load models into the test system one by one to evaluate the transfer limits under those models. The assessment results can provide a guideline to the system opearaters about the performances of each load model type. This evalution method can be extened to any other system typology.

The whole assessment process can be summarized into three steps. In the first step, we collect a set of reference P, Q dynamic responses from an unknown load model under one contingency. Then in the second step, we fit these reference responses into different load models through the introduced reinforcement learning load modeling method; in the last step, we run transfer limit assessment on all the fitted load models to evaluate their corresponding transfer limits.

## IV. CASE STUDY

The case study is conducted on the IEEE 39-bus system using Transient Security Assessment Tool (TSAT). We divided the IEEE 39-bus system into three areas, as shown in Fig.5. Transmission line 14-15 is the tie line between Area 1 and Area 2; transmission line 15-16 is the tie line between Area 2 and Area 3; transmission lines 1-39 and 3-17 are the tie lines between Area 1 and Area 3.

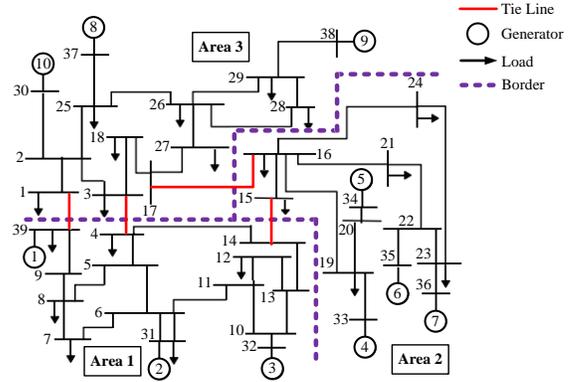

Figure 5. IEEE 39-bus system diagram

### A. Collect the Event Record

We first investigate the transfer limit between Area 2 and Area 3, in which the power sink is chosen to be the load on bus 20 in Area 2, and the power sources are generators 8, 9, and 10 on bus 37, 38 and 30 in Area 3. The reference load model uses WECC CLM with the load component composition as $[f_{Ma}, f_{Mb}, f_{Mc}, f_{Md}, f_{elc}, f_{zip}] = [0.1, 0.15, 0.1, 0.2, 0.1, 0.45]$. Due to the space limit, other parameters for this reference load model are omitted. When a three-phase to ground fault occurs at bus 6 in Area 1, we collect the system dynamic responses.

### B. Fit the Event Record Using Different Load Model

We first fitted the event data into the ZIP model, the best-fitted load composition is $[p_{1c}, p_{2c}, p_{3c}, q_{1c}, q_{2c}, q_{3c}] =$

$[\frac{2}{15}, \frac{1}{3}, \frac{8}{15}, \frac{3}{15}, \frac{1}{3}, \frac{7}{15}]$. The learning reward converged after around 3,500 episodes, as shown in Fig. 6a. The fitting result is shown in Fig. 6b. However, since ZIP does not have dynamic components, it cannot full replicate the event record dynamic responses. The P fitting RMSE is 0.0223 p.u., and the Q fitting RMSE is 0.0247 p.u.

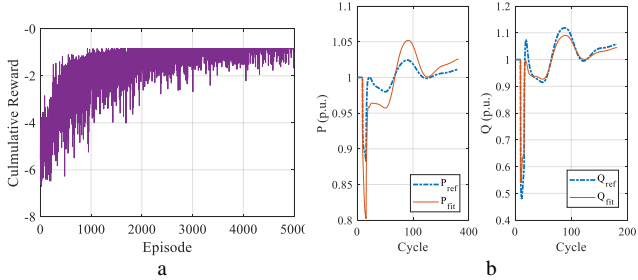

Figure 6. a. Trend of reward during the tranining process. b. Dynamic response compasison between the fitted ZIP model and field measurements

In addition to ZIP model, we also fit the field measurement into WECC CLM, ZIP + IM and several other reference models, including 50% ZIP + 50% IMa, 50% ZIP + 50% IMb, and 50% ZIP + 50% IMc. Due to space limits, the training processes to those models are not provided, but the P and Q fitting results for WECC CLM, and ZIP + IM are shown in Fig. 7 and Fig. 8. The fitting accuracies for P and Q of each load model are listed in Table I. The parameters' selection ranges for the IMs in ZIP +IM, 50% ZIP+ 50% IMA, 50% ZIP+ 50% IMB, and 50% ZIP+ 50% IMC models are different and defined based on the discussions made in [20].

Some empirical static model used in the industry are also included for evaluation, such as 40% Z + 60% P and 30% Z + 30% I + 40% P. Those static load models have fixed load compositions, therefore they can be directly implemented without a estimation process.

TABLE I
P&Q FITTING ACCURACY FOR THE SIX LOAD MODELS

| Load Model | WECC CLM | ZIP | ZIP + IM | 50% ZIP + 50% IMa | 50% ZIP + 50% IMb | 50% ZIP + 50% IMc |
|---|---|---|---|---|---|---|
| $RMSE_P$ | 0.0011 | 0.0223 | 0.0032 | 0.0227 | 0.0261 | 0.0254 |
| $RMSE_Q$ | 0.0046 | 0.0247 | 0.0190 | 0.0250 | 0.0175 | 0.0166 |

*All the number in Table I has unit of p.u.

Table I summarizes the estimation accuracy of each load model, and WECC CLM achieves the best modeling performance. Because with four dynamic components be included, the WECC CLM has a much higher degree of freedom to adjust its dynamics compared with other load models. The ZIP + IM is the second-best one. Because the induction motor model contained in the ZIP + IM model is a general model that can simulate the mixture of IMA, IMB, and IMC and makes it capable to fit the dynamic responses fairly well. Other models with a specific type of induction motor do not accurate enough to model the reference dynamic curves. The ZIP model has the worst fitting accuracy due to a lack of dynamic components.

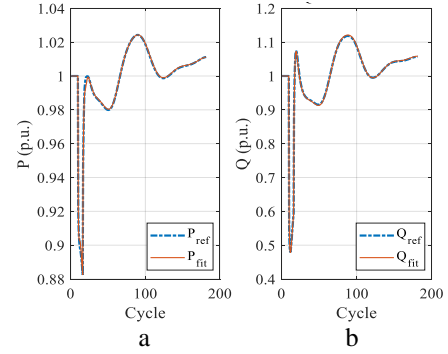

Figure 7. a. WECC CLM fiiting result on active power *P*. b. WECC CLM fiiting result on reactive power *Q*

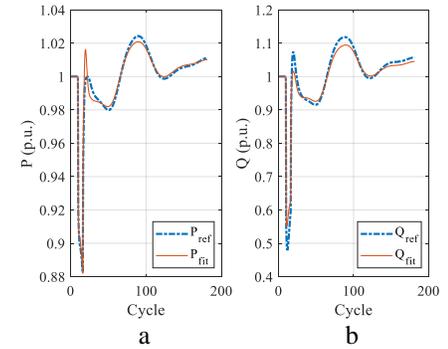

Figure 8. a. ZIP+IM fiiting result on active power *P*. b. ZIP fiiting result on reactive power *Q*

From the perspective of load modeling performance, the load models with more diverse dynamic components have higher flexibilities in approximating different dynamic responses, therefore, they can achieve better modeling accuracy. On the other hand, static load models have simple structures and are easy to fit, but their fitting accuracy is not satisfactory when dynamic features exist in the reference curves.

*C. Transfer Limit Comparison*

We individually swap the eight fitted load models shown in Table II onto the bus that designed to be the load-sink in transfer limit analysis. Then an N-1 contingency analysis is conducted for each load model. The transfer limits between any two zones with respect to different load models are repeatedly conducted and the results are shown in Table II.

We made the following observations on the transfer limits of the IEEE 39 Bus system from Table II: (1). the WECC CLM always leads to the lowest transfer limits; (2). 30% Z + 30% I + 40% P always provides the highest transfer limits; (3). ZIP + IM achieves lower transfer limits than ZIP but higher than the WECC CLM; (4) the transfer limits of 50% ZIP+ 50% IMA, 50% ZIP+ 50% IMB, and 50% ZIP+ 50% IMC models are higher than the ZIP +IM model but lower than the ZIP model. It can be concluded from those observations that the model complexity negatively affects the transfer limits in the system. With multiple motors modeled in the WECC CLM, it poses more dynamic constraints to the system which severely lower the transfer limits. However, the extent of impacts of load models on the transfer limit is case by case. For example, in Case I, the transfer limit of using WECC CLM is much less than

TABLE II
TRANSFER LIMIT (MW) COMPARISON FOR ALL 8 CASES

| Case | | Source | Sink | Tie Line | WECC CLM | ZIP | ZIP + IM | 50% ZIP + 50% IMa | 50% ZIP + 50%IMb | 50% ZIP + 50%IMc | 40% Z + 60% P | 30% Z + 30% I + 40% P |
|---|---|---|---|---|---|---|---|---|---|---|---|---|
| I | Area 2-3 | 10,8,9 | 20 | 16-17 | **1660** | 2272 | 2126 | 2204 | 2261 | 2244 | 2329 | **2361** |
| II | Area 2-3 | 10,8,9 | 21 | 16-17 | **2062** | 2333 | 2098 | 2140 | 2233 | 2200 | 2370 | **2380** |
| III | Area 1-2 | 4,5,6,7 | 4 | 14-15 | **2632** | 2824 | 2746 | 2774 | 2707 | 2660 | 2869 | **2878** |
| IV | Area 1-2 | 4,5,6,7 | 7 | 14-15 | **2667** | 2802 | 2751 | 2775 | 2735 | 2701 | 2821 | **2829** |
| V | Area 1-3 | 1,2,3 | 3 | 3-4, 1-39 | **1415** | 1451 | 1415 | 1423 | 1427 | 1426 | 1470 | **1477** |
| VI | Area 1-3 | 1,2,3 | 18 | 3-4, 1-39 | **1426** | 1464 | 1436 | 1442 | 1451 | 1443 | 1477 | **1483** |

*All the number under Source, Sink, and Tie Line indicates bus ID; all the transfer limit results has unit of MW.

other models, but the impacts are less significant in other cases.

According to experimental results, using static load models to simulate load dynamics can lead to a significantly higher transfer limit estimation than using dynamic load models. Among static load models, the ZIP model tends to less overestimate and reaches closer results to ZIP + IM. In addition, a static load model with fixed load component fractions is not adequate in modeling the constantly changing loads in the real world. In general, it is suggested to regularly update the load composition in any load models to better reflect the current load characteristics. From the transfer limit perspective, dynamic or composite load models can provide a secured or even conservative estimate, but the static load models tend to overestimate the transfer limits.

## V. CONCLUSION

In this paper, we evaluate different load models' impacts on the transfer limits using the IEEE 39-bus system. The reinforcement learning-based load modeling method is applied to estimate the model parameters with given historic event records. Through comprehensive case studies, we conclude that the dynamic load models, such as WECC CLM and ZIP + IM, have better capability to capture dynamic load behaviors at the cost of conservative transfer limits compared with static load models. The commonly used static load models cannot represent the complicated transient dynamics and it tends to overestimate the transfer limit. In addition, the load model complexity can negatively impact the transfer limit. The more complex the load model is, the less the calculated transfer limit will be. The proposed transfer limit evaluation method can also be applied to other systems.